\documentclass[pdflatex, sn-standardnature]{sn-jnl}% Standard Nature Portfolio Reference Style
\usepackage{amsthm,amsmath}
\usepackage{pdfpages}
\usepackage{enumerate}
\usepackage{tabularx}
\usepackage{lineno}
\jyear{2022}%

\theoremstyle{thmstyleone}%
\theoremstyle{thmstyletwo}%

\theoremstyle{thmstylethree}%

\begin{document}

\title[SOHPIE-DNA]{Differential Co-Abundance Network Analyses for Microbiome Data Adjusted for Clinical Covariates Using Jackknife Pseudo-Values}

%%=============================================================%%
%% Prefix	-> \pfx{Dr}
%% GivenName	-> \fnm{Joergen W.}
%% Particle	-> \spfx{van der} -> surname prefix
%% FamilyName	-> \sur{Ploeg}
%% Suffix	-> \sfx{IV}
%% NatureName	-> \tanm{Poet Laureate} -> Title after name
%% Degrees	-> \dgr{MSc, PhD}
%% \author*[1,2]{\pfx{Dr} \fnm{Joergen W.} \spfx{van der} \sur{Ploeg} \sfx{IV} \tanm{Poet Laureate} 
%%                 \dgr{MSc, PhD}}\email{iauthor@gmail.com}
%%=============================================================%%

\author[1]{\fnm{Seungjun} \sur{Ahn}}\email{sahn1@ufl.edu}

\author*[1]{\fnm{Somnath} \sur{Datta}}\email{somnath.datta@ufl.edu}

\affil[1]{\orgdiv{Department of Biostatistics}, \orgname{University of Florida}, \orgaddress{\street{2004 Mowry Road, 5th Floor CTRB}, \city{Gainesville}, \postcode{32611}, \state{FL}, \country{U.S.A}}}

%%==================================%%
%% sample for unstructured abstract %%
%%==================================%%

\abstract{A recent breakthrough in differential network (DN) analysis of microbiome data has been realized with the advent of next-generation sequencing technologies. The DN analysis disentangles the microbial co-abundance among taxa by comparing the network properties between two or more graphs under different biological conditions. However, the existing methods to the DN analysis for microbiome data do not adjust for other clinical differences between subjects. We propose a Statistical Approach via Pseudo-value Information and Estimation for Differential Network Analysis (SOHPIE-DNA) that incorporates additional covariates such as continuous age and categorical BMI. SOHPIE-DNA is a regression technique adopting jackknife pseudo-values that can be implemented readily for the analysis. We demonstrate through simulations that SOHPIE-DNA consistently reaches higher recall and F1-score, while maintaining similar precision and accuracy to existing methods (NetCoMi and MDiNE). Lastly, we apply SOHPIE-DNA on two real datasets from the American Gut Project and the Diet Exchange Study to showcase the utility. The analysis of the Diet Exchange Study is to showcase that SOHPIE-DNA can also be used to incorporate the temporal change of connectivity of taxa with the inclusion of additional covariates. As a result, our method has found taxa that are related to the prevention of intestinal inflammation and severity of fatigue in advanced metastatic cancer patients.}


%%\pacs[JEL Classification]{D8, H51}

%%\pacs[MSC Classification]{35A01, 65L10, 65L12, 65L20, 65L70}

\maketitle

\section{Introduction}
The human microbiome is the collective genomes of microbes or micro-organisms localized to the various sites of human body \cite{introMiNW1}. Recent clinical studies have shown that the microbiome has a regulatory role in a wide array of illnesses in humans, such as cancer \cite{MiNWCancer}, human immunodeficiency virus \cite{MiNWHIV}, and inflammatory bowel disease (IBD) \cite{MiNWIBD}. Moreover, the human microbiome is linked to emotional well-being \cite{MiNWemotional} and mental health including depression \cite{MiNWdepression}, autism spectrum disorders \cite{MiNWautism}, and human brain diseases \cite{MiNWbrain}. 

Following the advent of next-generation sequencing technologies, the taxonomic composition of microbial communities is better characterized by the amplification of small fragments (or amplicon) of the 16S ribosomal RNA (or 16S rRNA) gene. More recently, shotgun metagenomic sequencing has become an alternative for microbial community profiling \cite{shotgunMiNW}. Either sequencing platform typically employs similarity-based clustering algorithms to group 16S rRNA sequences into Operational Taxonomic Units (OTU) \cite{introMiNW2, introMiNW3} that are compositional.

The applications of network theory have been successfully utilized to better appraise the complex symbiotic (or dysbiotic) relationship between microbiome and disease states -- microbial co-abundances \cite{MiNWintroNW}. The abundance matrix or the observed OTU table is used to infer microbial co-abundances among taxa through either correlation-based approaches or probabilistic graphical models. 

The differential network (DN) analysis compares the network properties between two or more graphs under different biological conditions, such as degree centrality. Based on the recent review article \cite{matchadoreview}, there are two methods that are newly available to the DN analysis for microbiome data: Microbiome Differential Network Estimation (MDiNE) \cite{MDiNE} and Network Construction and comparison for Microbiome data (NetCoMi) \cite{NetCoMi}. These methods, however, do not incorporate additional covariates associated with the host or the composition of the microbiome.

It has been recognized that the composition of the gut microbiome is central to the pathogenesis of IBD \cite{MiNWIBD2, MiNWIBD}. In addition, the gut microbiome composition in patients with IBD is largely influenced by various factors including the use of antibiotics, diet, and cigarette smoking \cite{MiNWIBD}. In an analogous fashion, it is not unreasonable to speculate that the structure of the microbial networks can also vary depending on these factors. Thereby, there is a need for statistical methods for DN analysis that can include additional one predictor variables. 

One way to accomplish this goal is to use a regression technique based on pseudo-values, a component to calculate the bias-corrected estimator of leave-one-out jackknife resampling procedure \cite{pseudo.efron}. The pseudo-value technique was first postulated by Andersen and his colleagues \cite{pseudoref1, pseudoref2} in the context of multi-state survival models with right-censored data. Since then, it has been well studied in various disciplines of statistics including the interval-censored data \cite{MiNWpseudoref3, MiNWpseudoref2}, clustered data \cite{MiNWpseudoref6, pseudoref5}, and machine learning methods \cite{MiNWpseudoref4, MiNWpseudoref5}. 

The ultimate benefit of this technique is its straightforward inclusion of additional covariates in the generalized linear model \cite{MiNWpseudoref1}. An asymptotic linearity and consistency of pseudo-values given covariates are shown with the second-order von Mises expansion \cite{pseudoref4, MiNWpseudoref7}. The pseudo-values can then be used as the response variable in a regression model with the covariates \cite{MiNWpseudoref8}. Several studies reported that the type I error is well controlled at a nominal level of 0.05 while maintaining a high statistical power under the quasi-likelihood generalized linear mixed model \cite{MiNWpseudoref9} and generalized estimating equations framework \cite{pseudoref6, pseudoref5} for pseudo-value regression approach.

Hence, we propose a regression modeling method for DN analysis that regresses the jackknife pseudo-values calculated from a degree centrality of taxa in a microbial network to directly estimate the effects of predictors. In this approach, the grouping variable itself could also be included in the regression model along with additional clinical covariates while regressing the pseudo-values. We loosely refer to this as a ``multivariable setting'', whereas in ``univariable settings'' only the grouping variable is utilized in a DN analysis.

In the present study, we introduce \textbf{S}tatistical Appr\textbf{O}ac\textbf{H} via \textbf{P}seudo-value \textbf{I}nformation and \textbf{E}stimation for \textbf{D}ifferential \textbf{N}etwork \textbf{A}nalysis (SOHPIE-DNA) that can include covariate information in analyzing microbiome data. We firstly demonstrate the plausibility of the proposed method by comparing the model performances with MDiNE and NetCoMi through simulations under multivariable and univariable settings. Furthermore, the SOHPIE-DNA is applied to illustrate its clinical utility by examining real data from the American Gut Project \cite{amgut} and the Diet Exchange Study \cite{dietexchange} to identify DC taxa with presence of covariates. All statistical analyses are performed in \textsf{R} version 4.0.2 (\textsf{R} Foundation for Statistical Computing, Vienna, Austria).

\section{Methods}
\subsection{Compositional Correlation-Based Methods for Network Estimation}
The correlation is a useful proxy measure for identifying co-abundances or dependencies among taxa (or OTUs) in a microbial network. The Sparse Correlations for Compositional Data (SparCC) \cite{sparCC} estimates the pairwise correlations of the log-ratio transformed OTU abundances. Of note, a recent method, namely a Pseudo-value Regression Approach for Network Analysis (PRANA) \cite{PRANA}, operates on gene expression data only, which therefore does not use a correlation measure that preserves the compositional profiling.

The co-abundance among taxa is described by a covariance matrix  $T \in \mathbb{R}^{p \times p}$ where the non-diagonal elements $t_{jk}$ are expressed by
\begin{equation} \label{eq:3.1}
\begin{split}
t_{jk} & \equiv \text{Var} \left( \log \frac{u_{j}}{u_{k}} \right) \\
    & =  \text{Var} \left( \log u_{j} \right) + \text{Var} \left( \log u_{k} \right) -2 \text{Cov} \left( \log u_{j}, \log u_{k} \right) \\
    & = \sigma_{j}^{2} + \sigma_{k}^{2} - 2\rho_{jk}\sigma_{j}\sigma_{k},
\end{split}
\end{equation}
where $u_j$ and $u_k$ are the fraction of OTU abundances, $\sigma_{j}^{2}$ and $\sigma_{k}^{2}$ are the variances of the log-transformed abundances, and $\rho_{jk}$ is the correlation of taxa $j$ and $k$, respectively. Moreover, the variance $t_{jj}$ is approximated by 
\begin{equation} \label{eq:3.2}
t_{jj} \cong (p-1) \sigma_{j}^{2} + \sum_{k \neq j} \sigma_{k}^{2},
\end{equation}  
where $j, k \in \{1, \dots p\}$. Then the correlation can be estimated by solving equations \ref{eq:3.1} and \ref{eq:3.2}:
\begin{equation} \label{eq:3.3}
\hat{\rho}_{jk} = \frac{\hat{\sigma}_{j}^{2} + \hat{\sigma}_{k}^{2} - \hat{t}_{jk}} {2 \hat{\sigma}_{j} \hat{\sigma}_{k}},
\end{equation}  
where $\hat{\sigma}_{j}$, $\hat{\sigma}_{k}$, and $\hat{t}_{jk}$ are the sample estimates of $\sigma_{j}$, $\sigma_{k}$, and $t_{jk}$, respectively.

Furthermore, SparCC takes an iterative approach under the assumption (``sparsity of correlations'' as in the original paper) that a small number of strong correlations exists in a true network, which hinders the detection of spurious correlations among taxa. 

Besides SparCC, we have attempted to use other compositional correlation measures for our differential network analysis. See the Discussion section for further details.

\subsection{Pseudo-value Approach}
Consider undirected network estimated from $n$ individuals. It can then be represented by the $p \times p$ association matrix that encodes the pairwise correlations $\hat{\rho}_{jk}$ between a pair of taxa $j,k \in \{1, \dots , p\}$. The association matrix is symmetric ($\hat{\rho}_{jk} = $  $\hat{\rho}_{kj}$) where the non-diagonal entries are either non-zero (\textit{i.e.,} some association between two taxa) or zero (\textit{i.e.,} no association between two taxa). The diagonal entries are all equal to one, because the network is assumed that there is no self-loop (\textit{i.e.,} a node cannot redirect to itself).

The network centrality has been studied to measure the extent of biological or topological importance that a node has in a network \cite{centrality1, centrality2}. For each taxa $k$, the network centrality is calculated as the marginal sum of the association matrix.
\begin{equation*}
    \hat{\theta}_{k} = \sum_{j=1}^{p} \hat{\rho}_{jk},
\end{equation*}
where $k = 1, \dots , p$. 

The jackknife pseudo-values \cite{pseudo.efron} for the $i^\text{th}$ individual and $k^\text{th}$ taxon are defined by:
\begin{equation}
\tilde{\theta}_{ik} = n\hat{\theta}_{k} - (n-1)\hat{\theta}_{k(i)}, \label{eq:minw1}
\end{equation}
where $\hat{\theta}_{k(i)}$ is the marginal sum of a taxon calculated based on the re-estimated association matrix using the microbiome data eliminating the $i^\text{th}$ subject.

The computational cost of the re-estimation process is dependent on the sample size, as for each taxa $k$ requires $n$ such calculations with the data size of $n-1$. A solution to speed up the processing time is the use of parallel computing such as \textsf{mclapply} function in \textsf{parallel} \textsf{R} package.

Let $Z \in \{1, 2\}$ be a binary group indicator and denote $\mathcal{G}_1 = \{i: Z_{i} = 1 \}$ and $\mathcal{G}_2 = \{i: Z_{i} = 2 \}$. Each group has the same set of $p$ taxa, but group-specific sample size $n_{z} = \lvert \mathcal{G}_z \rvert$ for the two groups $z=1, 2$. Total sample size is $n = \sum_z n_{z}$. The equation \ref{eq:minw1} is used to calculate the group-specific jackknife pseudo-values. That is, for taxon $k$ and group $z$, we define $\hat{\theta}_{k}^z$ and $\hat{\theta}_{k(i)}^z$, where $i = 1,\ldots, n_{z}$. Then for each $i \in \mathcal{G}_z$, the $k^\text{th}$ taxon jackknife pseudo-values are calculated from $\tilde{\theta}_{ik} = n_z\hat{\theta}_{k}^z - (n_z-1)\hat{\theta}_{k(i)}^z$.

Let $\textbf{X} = (X_{1}, \dots , X_{q}$) denote $q$ vector of covariates, such as age at diagnosis, current smoking status, and etc. The pseudo-value regression model for the $i^\text{th}$ individual and $k^\text{th}$ taxon is
\begin{equation} 
%\begin{split}
\mu_{i} = E[\tilde{\theta}_{ik} \mid Z_{i}, \textbf{X}_{i}]
= \alpha_{k} + \beta_{k}Z_{i} +  \sum_{m=1}^{q} \gamma_{km} X_{im},
\label{eq:minw2}
%\end{split}
\end{equation}
where $\mu_{i}$ is the $k$-dimensional mean vector of pseudo-value $\tilde{\theta}_{ik}$ for the $i$th individual, $\alpha_{k}$ is the intercept, $\beta_{k}$ is the regression coefficient for $Z$, and $\gamma_{k1}, \dots , \gamma_{kq}$ is the set of regression coefficients to be estimated for $\textbf{X}$. In our setting, the main parameter of interest is given by $\beta_{k}$, the change in network centrality measure of the $k^{\text{th}}$ taxon between two groups.

The least trimmed squares (LTS), also known as least trimmed sum of squares \cite{LTS}, is then implemented to carry out a robust regression. The main advantages of the LTS estimator over other robust estimators including the M-estimator and least median of squares estimator are its computational efficiency and robustness to outliers in both the response and predictor variables \cite{LTS2, LTS3}.

The LTS estimator is defined by
\[
\min_{\alpha_k, \beta_k, \gamma_{k1}, \ldots, \gamma_{kq}} \sum_{i=1}^h r_{(i)} (\alpha_k, \beta_k, \gamma_{k1}, \ldots, \gamma_{kq})^2,
\]
where $r_{(i)}$ is the set of ordered absolute values of the residuals sorted in increasing order of absolute value and $h$ may depend on a pre-determined trimming proportion $c \in [0.5, 1]$ \cite{metrika}. For example, one can take $h = [n(1-c)] + 1$.

\subsection{Hypothesis Testing}
We construct the null hypothesis of $H_{0}: \beta_{k} = 0$ against the research hypothesis $H_{1}: \beta_{k} \neq 0$ to test if there is a true difference between groups in the network centrality measure of the $k^\text{th}$ taxon. The \textit{t}-statistic is defined by $U_{k} = \hat{\beta}_{k}/SE(\hat{\beta}_{k})$ for $k = 1, \dots, p$, where $\hat{\beta}_{k}$ is the least-squares estimator from the robust regression described in the above equations \ref{eq:minw2} and $SE(\hat{\beta}_{k})$ is the standard error of $\hat{\beta}_{k}$. As far as the decision-making process, the asymptotically $\alpha$-level test rejects $H_{0}$ if $\lvert U_{k} \rvert > t_{\alpha/2}$. P-values are calculated using a \textit{t}-distribution as in \textsf{robustbase} \textsf{R} package \cite{robustbase1, robustbase2}.

Multiple hypothesis testing is a common feature in the DN analysis, and therefore it is crucial to appropriately control the false discovery rate (FDR). The FDR measures the proportion of false discoveries incurred among a set of DC taxa from the test. Most classically, the concept of FDR was pioneered by Benjamini and Hochberg \cite{BH1}, shown to achieve the FDR control, whilst maintaining the adequate statistical power \cite{BH2}. However, the q-value \cite{qval2} offers a less conservative FDR estimation over the conventional Benjamini-Hochberg procedure \cite{qval1}. The q-value is estimated from the empirical distribution of the observed p-values, and keeps the balance between true positives and false positives \cite{qval3}. Accordingly, the q-value is applied to adjust for the multiplicity control in the present paper using \textsf{fdrtool} \textsf{R} package.

\subsection{Algorithm}

The SOHPIE-DNA algorithm is described below in Algorithm 1.
\begin{algorithm}
\small
\caption{SOHPIE-DNA}
\begin{algorithmic}[1]
\Require $n_{z} \times p$ OTU table and metadata for each group $z = 1, 2$.
\Ensure The set of p-values of the group variable for each taxa $k$.
\State Estimate $p$ $\times$ $p$ association matrix with SparCC from the $n_{z} \times p$ OTU table for each group $z = 1, 2$. See the previous section \ref{MiNW:Data Generation} for the data generation.
\State Calculate the group-specific marginal sums of association matrix of each taxa $k \in \{1, \cdots ,p\}$, denoted by $\hat{\theta}_{k}^z$.
\State Calculate $\hat{\theta}_{k(i)}^{z}$ for each taxa $k$ and individual $i \in \mathcal{G}_z$ from the association matrix that is re-estimated from the OTU table without the $i^{th}$ individual of $n_{z} \times p$ data.
\State Calculate the jackknife pseudo-value $\tilde{\theta}_{ik}$ using equation \ref{eq:minw1}.
\State Fit a robust regression for each taxa $k$ to obtain the p-values of the group variable, computed from the $t$-test.
\begin{enumerate}[(i)]
     \item Multivariable: A binary group variable $Z$ and a continuous covariate $X$ are included in the model.
     \item Univariable: A binary group $Z$ is only included in the model.
\end{enumerate}
\State The q-values are calculated based on the observed p-values for the multiplicity control. This will be used to compute the performance measures of the Monte Carlo simulation (see the next section \ref{MiNW:measures}).
\end{algorithmic}
\end{algorithm}

\subsection{Performance Evaluations}
\subsubsection{Construction of Adjacency Matrices}
Generate the scale-free random network (or Barab\'{a}si-Albert network) \cite{barabasi} with $p$ nodes using the \textsf{igraph} \textsf{R} package \cite{igraphR}. A network is scale-free if its degree distribution follows a power-law distribution. In other words, a small portion of ``hub'' nodes has the highest degree centrality, while most nodes have lower degree centrality.

\begin{figure}[hbt!]
\centerline{\includegraphics[scale=0.3]{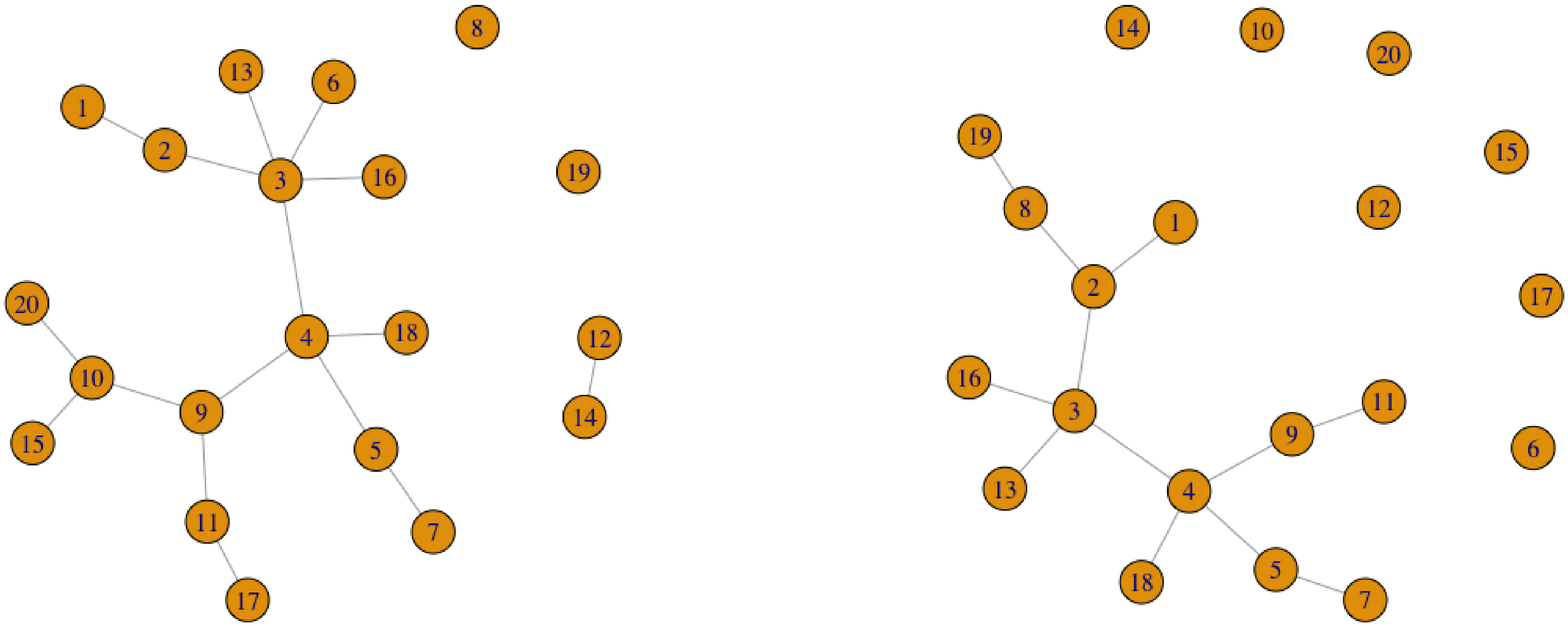}}
\caption{Network plots visualizing the microbial network ($p = 20$) with a covariate dependence structure that depends on continuous age and binary group information ($\delta_{1} = 0.05$ (left), $\delta_{2} = 0.2$ (right)). This represents the multivariable setting.}
\label{MiNW:Fig1}
\end{figure}

\begin{figure}[hbt!]
\centerline{\includegraphics[scale=0.3]{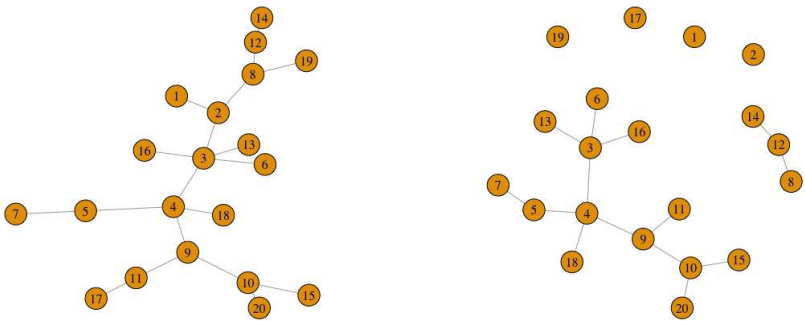}}
\caption{Network plots visualizing the microbial network ($p$ = 20) without a covariate dependence structure that depends on binary group only ($\delta_{1} = 0$ (left), $\delta_{2} = 0.2$ (right)). This represents the univariable setting.}
\label{MiNW:Fig2}
\end{figure}

The two identical $p \times p$ adjacency matrices, where the diagonal entries are 0 and non-diagonal entries are either $\{0, 1\}$, are obtained from this random network. At the end of the data generation phase using \textsf{SparseDOSSA2} in Simulated Data section \ref{MiNW:Data Generation}, we are able to identify which taxa are spike-in associated with the covariate for each $z = 1, 2$. In order to distinguish networks representative of $z = 1$ (\textit{e.g.,} healthy control) from that of $z = 2$ (\textit{e.g.,} disease group), we keep track of the indices of these covariate-dependent taxa. We perturb the random networks by removing all the connected edges around nodes of the adjacency matrix using the indices recorded previously for each group. The network plots are provided to visually demonstrate the perturbed adjacency matrices (see Figure \ref{MiNW:Fig1} and \ref{MiNW:Fig2}).

\subsubsection{Performance Measures}
\label{MiNW:measures}
Four performance metrics are adopted to evaluate our proposed method: precision, recall, F1-score, and accuracy. Let $\Omega^{z} \in \mathbb{R}^{p \times p}$ be the group-specific adjacency matrix, where

\begin{equation*}
    \Omega_{jk}^{z} =
        \begin{cases}
          1 & \text{if the two nodes } j \text{ and } k \text{ are connected}\\
          0 & \text{otherwise},
        \end{cases}
\end{equation*}

for $z = 1, 2$. Next, a node-specific true connection is calculated
\begin{equation*}
    \eta_{k} = I\bigg( \sum_{j=1}^{p} \lvert \Omega_{jk}^{1} - \Omega_{jk}^{2} \rvert > 0 \bigg),
\end{equation*}
indicating that taxa $k$ has differential connectivity (DC).

In terms of notation, we use $q_{ks}$ to denote a q-value \cite{qval2} of taxa $k$ at the simulation replicate $s$. An error rate control of $\alpha = 0.05$ is used throughout the simulation. In the following, we present the details of each performance metric.

Precision is the fraction of taxa which are declared to be significantly DC from the test that are confirmed as true:
\begin{equation*}
      \text{Precision} = \frac{ \sum_{k = 1}^{p} \eta_{k} \, I(q_{ks} < \alpha) }{\sum_{k = 1}^{p} I(q_{ks} < \alpha) }.
  \end{equation*}
  
Recall is the fraction of truly DC taxa which are correctly declared to be significant between two comparing groups from the test:
\begin{equation*}
      \text{Recall} = \frac{ \sum_{k = 1}^{p} \eta_{k} \, I(q_{ks} < \alpha) }{\sum_{k = 1}^{p} \eta_{k} }.
  \end{equation*}
%\section*{Section title}
%Text for this section\ldots
%\subsection*{Sub-heading for section}
%Text for this sub-heading\ldots

%\cite{koon,xjon,marg,schn,koha,issnic}).

The F1 score is the harmonic mean of precision and recall values. A higher F1 score indicates a better overall performance with lower false negative and false positive predictions:
\small
  \begin{equation*}
      \text{F1} = 2 \cdot \frac{\text{Precision} \cdot \text{Recall}}{\text{Precision} + \text{Recall}}.
  \end{equation*}
\normalsize
Accuracy is defined as the fraction of total number of taxa that are correctly predicted to be DC. The accuracy ranges from 0 (no correct predictions) to 1 (perfect predictions):
\small
  \begin{equation*}
      \text{Accuracy} = \frac{ \sum_{k = 1}^{p} I(\eta_{k}=1) \, I(q_{ks} < \alpha) + \sum_{k = 1}^{p} I(\eta_{k}=0) \, I(q_{ks} \geq \alpha) }{\sum_{k = 1}^{p} I(\eta_{k}=1) + \sum_{k = 1}^{p} I(\eta_{k}=0) }.
  \end{equation*}
\normalsize

\subsection{Materials}
\subsubsection{Simulated Data}\label{MiNW:Data Generation}
The synthetic microbiome dataset are structured with $p$ taxa and $n$ sample size. In the simulation, binary group indicators 1 and 2 are generated from a Bernoulli distribution with equal probabilities and a single continuous covariate $X \sim N(55, 10)$ (\textit{e.g.,} age at diagnosis). We test our proposed method on datasets under two different simulation scenarios: taxa are impacted by the effect of (1) $Z$ and $X$ or (2) $Z$ only, which each corresponds to ``multivariable'' and ``univariable'' settings, respectively. 

The actual microbial data generation (\textit{e.g.,} OTU counts) given the covariates is described next. In this context, it is perhaps worth mentioning that this part is completely different from generating gene expression data as in PRANA \cite{PRANA}. For each simulation scenario, we generate an OTU table that resembles the dependence structure of covariates $Z$ and/or $X$ on the microbial community (or the network) using the \textsf{SparseDOSSA2} (Sparse Data Observations for the Simulation of Synthetic Abundances) \textsf{R} package \cite{SparseDOSSA2}. \textsf{SparseDOSSA2} adopts a Bayesian Gaussian copula model with zero-inflated, truncated log-normal distributions to capture the marginal distributions of each microbial taxa and to account for the correlation between taxa.

The package has a feature to indicate a user-specified percentage of taxa to be ``spiked-in'' association with the clinical information (or metadata). This is referred to as the ``effect size'' of differential abundance $\delta$. To evaluate the effect size of $Z$ under the univariable setting, we generate the data that half of the samples have taxa with no spike-in association, whereas the other half of the samples have spike-in association on $5\%$, $10\%$, or $20\%$ of taxa. The distributions of age in the two groups are different. Therefore, under the multivariable setting, $5\%$, $10\%$, or $20\%$ of taxa have spike-in association with $X$ for each group $z = 1, 2$. In both scenarios, $n_{z} \times p$ matrices for each group $z = 1, 2$ will be available for use.

\subsubsection{Application Study}
\paragraph{The American Gut Project Data}
A pre-processed OTU table of the human stool microbiome samples from the American Gut Project \cite{amgut} is available in the \textsf{SpiecEasi} \textsf{R} package, along with the corresponding metadata information. The gut microbiome is involved with the bidirectional relationship between the gastrointestinal system and central nervous system (\textit{i.e.} gut-brain axis) that impacts on the migraine inflammation \cite{migraine_gut}. 

In the analysis, the main variable of interest is a binary variable indicating the migraine headache (yes or no). Age \cite{migraine_agesex}, sex \cite{migraine_agesex}, exercise frequency ($\geq 3$ days per week or otherwise) \cite{migraine_exercise}, and categorical alcohol consumption (heavy, moderate, or non-drinking) \cite{migraine_alcohol} are covariates that are included in the multivariable model. Additionally, migraine has been associated with the periodontal inflammation \cite{migraine_dental} and pet ownership \cite{migraine_pet}, and therefore the oral hygiene behavior such as dental floss frequency ($\geq 3$ times per week or otherwise) and living with a dog (yes or no) were included in the model.

The initial OTU table consists of 138 taxa with 296 subjects. No taxa were removed, however, 28 subjects were excluded due to unidentified sampling body site and missing age or sex information. Hence, 138 taxa and 268 subjects were used for the analysis.

\paragraph{The Diet Exchange Study Data}
A pre-processed data of the geographical epidemiology study \cite{dietexchange} is available in \textsf{microbiome} \cite{microbiomeR} \textsf{R} package. The aim of the study was to assess the effect of fat and fiber intake of the diet on the composition of the colonic microbiota by switching the diet in study populations with high (African-Americans from Pittsburgh area of Pennsylvania; AA) and low (rural South Africans from KwaZulu region; RA) colon cancer risk for two weeks. 

An initial OTU table contains 130 taxa with 38 subjects. After the exclusion of a subject with missing post-dietary intervention data and 18 rare taxa that appear in fewer than 10$\%$ of the samples, 112 taxa with 37 subjects (20 AA and 17 RA) are used for the analysis. 

The main predictor variable is binary geographic location (AA or RA). Additional covariates considered in a multivariable model were sex and BMI groups (obese, overweight, or lean).

For each groups separately, we take the difference of the estimated association matrices (as well as the re-estimated association matrices) between two time points, that is, the endoscopy before and after two weeks of dietary change. The differences are then used to calculate the jackknife pseudo-values as in the previous sections. This additional step is intended to incorporate the temporal change of connectivity of each taxa after dietary interventions.

\section{Results}
\subsection{Simulation Study}
The sample size $n = 20, 50, 200, 500$ are considered for each microbial network with $p = 20, 40$ taxa over 1,000 Monte Carlo replicates. Simulations are repeated to assess the effects of covariates on taxa by changing the effect size, $\delta = 0.05, 0.1, 0.2$, which is described in Simulated Data section \ref{MiNW:Data Generation}. A new network is generated at each simulation replicate to account for biological variability of the network structure. 

The performance metrics provided in section \ref{MiNW:measures} are computed by comparing the test results with the true network. In the true network setting, a taxa is truly DC between groups if it is connected to at least one neighbor taxa. Tables 1 and 2 summarize simulation results under the multivariable setting. That is, a continuous covariate is included with the binary group variable in the regression model. To illustrate the utility of the proposed method on covariate-dependent network, we compared the pseudo-value regression approach with the recent methods available (NetCoMi and MDiNE) that cannot incorporate the additional covariate. Results show that the SOHPIE-DNA consistently maintains high recall values in all specifications of taxa, sample sizes, and effect sizes, and outperforms NetCoMi and MDiNE in almost all cases. A higher F1 score of SOHPIE-DNA indicates that the proposed method can achieve a better overall model performance in the presence of additional covariates, compared with the two competing methods. In general, all metrics improve as $n$ increases and/or when the larger effect size is provided ($\delta=0.2)$, as expected. It is worth noting that the MDiNE poses a practical challenge associated with substantially large computational time and costs. For instance, it requires more than 9 days to complete each simulation for $p=40$ and $n=200$ from the University of Florida Research Computing Linux server, HiPerGator 3.0 with 32CPU cores and 4GB of RAM per node, while it takes up to 18 hours to execute the same simulation tasks for both the SOHPIE-DNA and NetCoMi with 4CPU cores and 6GB of RAM per node. See Table S1 in Additional File 1 for more details.

Table 3 presents results of the univariable setting, where only the binary group variable is included in the model. In other words, only the effect of group was considered when generating random networks. On the whole, a similar pattern is shown in the univariable setting that the SOHPIE-DNA reaches a higher level of recall, compared with NetCoMi and MDiNE. Overall, our method resulted in a higher F1 score when the smaller network is considered. All of the methods suffer from a low precision with a small effect size ($\delta=0.05)$, but eventually improves with a larger effect size ($\delta=0.2)$.

\subsection{Analysis of the American Gut Project Data}
Six out of 138 taxa are found significantly DC between migraineurs vs. non-migraineurs while adjusting for age, sex, exercise frequency, categorical alcohol consumption, oral hygiene behavior, and dog ownership. At the family-level, the DC taxa are members of \textit{Ruminococcaceae}, \textit{Lachnospiraceae}, \textit{Enterobacteriaceae}, \textit{Erysipelotri-chaceae}, and \textit{Bacteroidaceae}. Of these families, the absence of \textit{Lachnospiraceae} has been linked to the active or severe \textit{Clostridium difficile} infection \cite{lachno1}. \textit{Erysipelotrichaceae} has been associated with dyslipidemic phenotypes and systemic inflammation \cite{erys1}. Moreover, a recent study \cite{lachno2} reported that the species enriched among migraineurs include \textit{Ruminococcus gnavus} and \textit{Lachnospiraceae bacterium}. The computational time for our analysis was about 12 hours on the high-performance Linux cluster, HiPerGator 3.0 with 16CPU cores and 4GB of RAM per node.

\subsection{Analysis of the Diet Exchange Study Data}
Out of 112 taxa, 16 are predicted to be significantly DC between AA and RA after the two-week dietary exchange intervention while accounting for their age and BMI group. A complete list of DC taxa represent \textit{Bacillus}, \textit{Bacteroides uniformis et rel.}, \textit{Bacteroides vulgatus et rel.}, \textit{Clostridium ramosum et rel.}, \textit{Coprococcus eutactus et rel.}, \textit{Eggerthella lenta et rel.}, \textit{Escherichia coli et rel.}, \textit{Eubacterium hallii et rel.}, \textit{Eubacterium siraeum et rel.}, \textit{Faecalibacterium prausnitzii et rel.}, \textit{Prevotella oralis et rel.}, \textit{Roseburia intestinalis et rel.}, \textit{Ruminococcus gnavus et rel.}, \textit{Staphylococcus}, \textit{Uncultured Bacteroidetes}, and \textit{Xanthomonadaceae}. Notably, \textit{Roseburia intestinalis} contributes to the prevention and management of intestinal inflammation and atherosclerosis \cite{diet_roseburia}. \textit{Eubacterium hallii} has been negatively associated with the fatigue severity scores of patients with advanced metastatic cancer \cite{diet_eubacterium}. The analysis took about an hour and 11 minutes on the HiPerGator 3.0 with 16CPU cores and 4GB of RAM per node.

\section{Discussion}
In this manuscript, we introduce the SOHPIE-DNA, a pseudo-value regression approach that determines whether a microbial taxa is significantly DC between groups after adjusting for additional covariates. This study is the first of its kind in the literature to develop a regression modeling for the DN analysis in microbiome data, which includes more than one predictor (\textit{e.g.,} group) in the model and predicts features of connectivity of a network. A simulation study shows that, at least for the scenarios considered, the SOHPIE-DNA generally maintains higher recall and F1-score while maintaining similar precision and accuracy, when compared with the most recent state-of-the-art methods: NetCoMi and MDiNE. 

In this study, the group-specific jackknife pseudo-values are calculated. Another way of calculating jackknife pseudo-values is to use the entire sample and use the group-level indicator as a covariate into the model. However, in our preliminary simulations, we found that doing it that way led to worse performance.

We analyzed the data from two published studies to showcase the utility of the SOHPIE-DNA. Firstly, 6 taxa are found to be significantly DC between migraineurs and non-migraineurs while accounting for covariates using the data from the American Gut Project. A slight modification to the proposed method is grafted for analyzing the Diet Exchange Study data, where the group-specific difference of the estimated association matrices between two time points are used for the pseudo-value calculation. As a result, 16 significantly DC taxa are identified between AA and RA after the two-week diet exchange intervention with the inclusion of covariates. 

The latter application demonstrates the capability of assessing the temporal variation in connectivity measures. However, the SOHPIE-DNA currently has no feature to address the within-subject correlation for repeated measurements at different time points. This opens up an avenue for future investigation of longitudinal microbiome studies. One way of handling this is to use a generalized estimating equations (GEE) type approach for the pseudo-values and utilizing a jackknife estimate of the variance-covariance matrix of the pseudo-values at different time points.

Another line of future research direction to extend our work is to consider the idea of variable selection. This will help finding the best prediction model with a subset of phenotypic variables that are more biologically relevant across more heterogeneous study samples.

Additionally, we made an attempt of fitting a model under the generalized linear model for binary outcomes: logistic regression with or without the Firth's correction, in case of small sample size. It was challenging to appropriately dichotomize the matrices with jackknife pseudo-values. Further studies will be needed to devise an adaptive algorithm to find a threshold value that better classify the jackknife pseudo-values.
 
As a last remark, it should be emphasized that methods other than SparCC were also considered for network estimation, which includes the CCLasso \cite{CCLasso} and SPIEC-EASI \cite{SpiecEasi} with graphical lasso or neighborhood selection algorithms. However, these were not favorable in terms of runtime or due to not being able to run under certain simulation scenarios. For instance, the computational time to complete the re-estimation step for the SPIEC-EASI took more than 200 minutes for $p = 20$ with $n = 200$ for a single simulation replicate. The CCLasso could not estimate the association matrix with small sample size for a smaller network ($p = 20$ for $n = 20, 40, 60$).

\section{Conclusion}
There has been limited research to date that discusses how to adjust for additional covariate information in DN analysis for microbiome data. Herewith, we propose SOHPIE-DNA, a novel pseudo-value regression approach for the DN analysis, which can include additional clinical covariate in the model.
\backmatter

%\bmhead{Supplementary information}

%If your article has accompanying supplementary file/s please state so here. 

%Authors reporting data from electrophoretic gels and blots should supply the full unprocessed scans for key as part of their Supplementary information. This may be requested by the editorial team/s if it is missing.

%Please refer to Journal-level guidance for any specific requirements.

\subsection*{Acknowledgments}
The authors are exceptionally thankful for the investigators involved with the American Gut Project and the Diet Exchange Study for sharing their data publicly. S.A. has been supported by Award Number [NIH T32AA025877] from the National Institute on Alcohol Abuse and Alcoholism of the National Institutes of Health. S.A. dedicates this work to remember all the memories that he had with his furry friend, Sofie. 

\section*{Declarations}
\subsection*{Author contributions}
S.D. conceived the use of pseudo-value in the study. S.A. developed the methodology, performed simulations and data analyses of the study. S.A. drafted the manuscript and S.D. provided suggestions when writing the manuscript. All authors have reviewed and edited the manuscript.
\subsection*{Competing interests}
The authors have no competing interests to disclose.
\subsection*{Data availability}
A pre-processed OTU table and metadata from the American Gut Project and from the Diet Exchange Study are available in the \textsf{SpiecEasi R} package and \textsf{microbiome R} package, respectively. Please reach out to the author (Somnath Datta, somnath.datta@ufl.edu) if you have any further inquiries.
\subsection*{Code availability}
SOHPIE-DNA is available at \url{https://github.com/sjahnn/SOHPIE-DNA}.
\subsection*{Funding}
S.A. is funded by the National Institute on Alcohol Abuse and Alcoholism at the National Institutes of Health under Award Number [NIH T32AA025877].
\subsection*{Ethics approval} 
Not applicable.
\subsection*{Consent to participate}
Not applicable.
\subsection*{Consent for publication}
Not applicable.

\noindent

\bibliography{sn-bibliography}% common bib file
%% if required, the content of .bbl file can be included here once bbl is generated
%%\input sn-article.bbl

%% Default %%
%%\input sn-sample-bib.tex%

\newpage

% Table 1 below (Multivariable results for p=20)
\begin{sidewaystable}
\centering
\caption{\label{MiNW:table1}The simulation results for the case when the network structure depends on age covariate. The binary group variable in the multivariable regression model (continuous age and binary group) using pseudo-value approach is compared with NetCoMi and MDiNE with 1,000 replicates. A random network with network size $p$=20 is generated at each simulation replicate. The best results are highlighted in boldface.}
\makebox[\textwidth][c]{
{\footnotesize%fontsize{5}{7}\selectfont
\begin{tabular}{@{}m{.2cm}m{.2cm}m{.3cm}m{.3cm} | m{1.0cm}m{1.1cm}m{1.1cm} |  m{1.0cm}m{1.1cm}m{1.1cm} |  m{1.0cm}m{1.1cm}m{1.1cm} | m{1.0cm}m{1.1cm}m{0.7cm} }
 \hline
 &  &  &  & \multicolumn{3}{c|}{Precision} & \multicolumn{3}{c|}{Recall} & \multicolumn{3}{c|}{F1} & \multicolumn{3}{c}{Accuracy} \\
\cline{5-7}\cline{8-10}\cline{11-13}\cline{14-16}
$p$ & $n$ & $\delta_{1}$ & $\delta_{2}$ & SOHPIE & NetCoMi & MDiNE & SOHPIE & NetCoMi & MDiNE & SOHPIE & NetCoMi & MDiNE & SOHPIE & NetCoMi & MDiNE \\
 \hline
20 & 20 & 0.05 & 0.05 & 0.26 & 0.28 & \textbf{0.36} & \textbf{0.69} & 0.25 & 0.01 & \textbf{0.39} & 0.35 & 0.31 & 0.42 & 0.06 & \textbf{0.75} \\
& & 0.05 & 0.10 & 0.36 & 0.37 & \textbf{0.49} & \textbf{0.69} & 0.26 & 0.01 & \textbf{0.45} & 0.37 & 0.30 & 0.45 & 0.09 & \textbf{0.65} \\
& & 0.05 & 0.20 & \textbf{0.52} & 0.51 & 0.39 & \textbf{0.69} & 0.24 & 0.01 & \textbf{0.57} & 0.36 & 0.24 & \textbf{0.51} & 0.12 & 0.49   \\ 
& & 0.10 & 0.05 & 0.36 & 0.36 & \textbf{0.44} & \textbf{0.70} & 0.23 & 0.01 & \textbf{0.46} & 0.35 & 0.32 & 0.45 & 0.08 & \textbf{0.65} \\
& & 0.10 & 0.10 & \textbf{0.42} & \textbf{0.42} & \textbf{0.42} & \textbf{0.69} & 0.25 & 0.01 & \textbf{0.51} & 0.37 & 0.25 & 0.47 & 0.11 & \textbf{0.58} \\
& & 0.10 & 0.20 & \textbf{0.54} & \textbf{0.54} & 0.43 & \textbf{0.70} & 0.24 & 0.01 & \textbf{0.59} & 0.38 & 0.28 & \textbf{0.52} & 0.13 & 0.46 \\
& & 0.20 & 0.05 & 0.50 & \textbf{0.51} & 0.48 & \textbf{0.69} & 0.25 & 0.01 & \textbf{0.56} & 0.38 & 0.25 & \textbf{0.51} & 0.12 & 0.50 \\
& & 0.20 & 0.10 & 0.53 & \textbf{0.54} & 0.45 & \textbf{0.70} & 0.26 & 0.01 & \textbf{0.58} & 0.38 & 0.20 & \textbf{0.52} & 0.14 & 0.47\\
& & 0.20 & 0.20 & \textbf{0.60} & 0.58 & 0.42 & \textbf{0.70} & 0.24 & 0.01 & \textbf{0.62} & 0.38 & 0.18 & \textbf{0.54} & 0.14 & 0.41   \\ 
& 50 & 0.05 & 0.05 & 0.25 & 0.27 & \textbf{0.35} & \textbf{0.82} & 0.53 & 0.12 & \textbf{0.39} & 0.37 & 0.34 & 0.34 & 0.13 & \textbf{0.72}   \\
& & 0.05 & 0.10 & 0.35 & 0.37 & \textbf{0.42} & \textbf{0.84} & 0.58 & 0.10 & \textbf{0.49} & 0.44 & 0.30 & 0.41 & 0.20 & \textbf{0.63} \\
& & 0.05 & 0.20 & 0.50 & \textbf{0.52} & 0.50 & \textbf{0.84} & 0.63 & 0.10 & \textbf{0.61} & 0.54 & 0.26 & \textbf{0.50} & 0.31 & \textbf{0.50}   \\
& & 0.10 & 0.05 & 0.35 & 0.37 & \textbf{0.42} & \textbf{0.83} & 0.58 & 0.11 & \textbf{0.49} & 0.44 & 0.30 & 0.40 & 0.20 & \textbf{0.63}\\
& & 0.10 & 0.10 & 0.41 & 0.43 & \textbf{0.44} & \textbf{0.83} & 0.62 & 0.11 & \textbf{0.54} & 0.48 & 0.29 & 0.44 & 0.25 & \textbf{0.57} \\
& & 0.10 & 0.20 & 0.53 & 0.55 & \textbf{0.57} & \textbf{0.84} & 0.64 & 0.12 & \textbf{0.64} & 0.56 & 0.28 & \textbf{0.52} & 0.34 & 0.47 \\
& & 0.20 & 0.05 & 0.51 & \textbf{0.52} & 0.49 & \textbf{0.84} & 0.61 & 0.10 & \textbf{0.62} & 0.54 & 0.27 & \textbf{0.51} & 0.31 & 0.49  \\
& & 0.20 & 0.10 & 0.54 & \textbf{0.55} & 0.54 & \textbf{0.83} & 0.64 & 0.11 & \textbf{0.64} & 0.57 & 0.27 & \textbf{0.52} & 0.35 & 0.47 \\
& & 0.20 & 0.20 & \textbf{0.59} & \textbf{0.59} & 0.58 & \textbf{0.84} & 0.69 & 0.12 & \textbf{0.68} & 0.61 & 0.27 & \textbf{0.56} & 0.40 & 0.43  \\
& 200 & 0.05 & 0.05 & 0.26 & \textbf{0.27} & 0.26 & \textbf{0.93} & 0.63 & 0.45 & \textbf{0.41} & 0.38 & 0.35 & 0.29 & 0.16 & \textbf{0.52} \\
& & 0.05 & 0.10 & 0.35 & \textbf{0.37} & 0.35 & \textbf{0.94} & 0.68 & 0.45 & \textbf{0.50} & 0.46 & 0.38 & 0.37 & 0.24 & \textbf{0.51} \\
& & 0.05 & 0.20 & 0.51 & \textbf{0.52} & 0.50 & \textbf{0.94} & 0.74 & 0.48 & \textbf{0.65} & 0.58 & 0.47 & \textbf{0.51} & 0.37 & 0.49   \\
& & 0.10 & 0.05 & 0.35 & \textbf{0.36} & 0.35 & \textbf{0.94} & 0.67 & 0.46 & \textbf{0.50} & 0.45 & 0.40 & 0.37 & 0.23 & \textbf{0.51} \\
& & 0.10 & 0.10 & 0.41 & \textbf{0.43} & 0.40 & \textbf{0.94} & 0.74 & 0.48 & \textbf{0.56} & 0.52 & 0.43 & 0.42 & 0.30 & \textbf{0.50} \\
& & 0.10 & 0.20 & 0.54 & \textbf{0.55} & 0.53 & \textbf{0.93} & 0.78 & 0.50 & \textbf{0.67} & 0.62 & 0.49 & \textbf{0.53} & 0.42 & 0.50 \\
& & 0.20 & 0.05 & 0.50 & \textbf{0.52} & 0.48 & \textbf{0.94} & 0.72 & 0.48 & \textbf{0.64} & 0.58 & 0.46 & \textbf{0.50} & 0.36 & 0.49 \\
& & 0.20 & 0.10 & 0.53 & \textbf{0.54} & \textbf{0.54} & \textbf{0.93} & 0.76 & 0.51 & \textbf{0.67} & 0.61 & 0.51 & \textbf{0.53} & 0.41 & 0.51\\
& & 0.20 & 0.20 & 0.58 & \textbf{0.59} & 0.56 & \textbf{0.93} & 0.81 & 0.52 & \textbf{0.71} & 0.66 & 0.52 & \textbf{0.57} & 0.48 & 0.49  \\ 
& 500 & 0.05 & 0.05 & 0.25 & \textbf{0.27} & 0.24 & \textbf{0.96} & 0.73 & 0.69 & \textbf{0.41} & 0.40 & 0.36 & 0.27 & 0.18 & \textbf{0.38}  \\
& & 0.05 & 0.10 & 0.35 & \textbf{0.37} & 0.35 & \textbf{0.97} & 0.78 & 0.75 & \textbf{0.51} & 0.48 & 0.47 & 0.36 & 0.28 & \textbf{0.42}\\
& & 0.05 & 0.20 & 0.51 & \textbf{0.52} & 0.48 & \textbf{0.96} & 0.82 & 0.76 & \textbf{0.65} & 0.61 & 0.57 & \textbf{0.51} & 0.42 & 0.48   \\
& & 0.10 & 0.05 & 0.34 & \textbf{0.35} & \textbf{0.35} & \textbf{0.96} & 0.76 & 0.71 & \textbf{0.50} & 0.46 & 0.45 & 0.35 & 0.26 & \textbf{0.43} \\
& & 0.10 & 0.10 & 0.42 & \textbf{0.43} & 0.42 & \textbf{0.97} & 0.80 & 0.74 & \textbf{0.57} & 0.53 & 0.52 & 0.42 & 0.33 & \textbf{0.46} \\
& & 0.10 & 0.20 & 0.53 & \textbf{0.54} & 0.51 & \textbf{0.96} & 0.85 & 0.79 & \textbf{0.67} & 0.64 & 0.60 & \textbf{0.53} & 0.45 & 0.50 \\
& & 0.20 & 0.05 & 0.50 & \textbf{0.51} & 0.49 & \textbf{0.96} & 0.82 & 0.78 & \textbf{0.65} & 0.61 & 0.59 & \textbf{0.50} & 0.41 & \textbf{0.50}  \\
& & 0.20 & 0.10 & 0.53 & \textbf{0.54} & 0.52 & \textbf{0.97} & 0.86 & 0.76 & \textbf{0.67} & 0.64 & 0.60 & \textbf{0.53} & 0.46 & 0.51\\
& & 0.20 & 0.20 & \textbf{0.59} & \textbf{0.59} & 0.58 & \textbf{0.96} & 0.88 & 0.83 & \textbf{0.72} & 0.68 & 0.66 & \textbf{0.58} & 0.51 & 0.56  \\ \hline
\end{tabular}}{}
}
\end{sidewaystable}
\bigskip

% Table 2 below (Multivariable results for p=40)
\begin{sidewaystable}
\centering
\caption{\label{MiNW:table2}The simulation results for the case when the network structure depends on age covariate. The binary group variable in the multivariable regression model (continuous age and binary group) using pseudo-value approach is compared with NetCoMi and MDiNE with 1,000 replicates. A random network with network size $p$=40 is generated at each simulation replicate. The best results are highlighted in boldface.}
\makebox[\textwidth][c]{
{\footnotesize%fontsize{5}{7}\selectfont
\begin{tabular}{@{}m{.2cm}m{.2cm}m{.3cm}m{.3cm} | m{1.0cm}m{1.1cm}m{1.1cm} |  m{1.0cm}m{1.1cm}m{1.1cm} |  m{1.0cm}m{1.1cm}m{1.1cm} | m{1.0cm}m{1.1cm}m{0.7cm} }
 \hline
 &  &  &  & \multicolumn{3}{c|}{Precision} & \multicolumn{3}{c|}{Recall} & \multicolumn{3}{c|}{F1} & \multicolumn{3}{c}{Accuracy} \\
\cline{5-7}\cline{8-10}\cline{11-13}\cline{14-16}
$p$ & $n$ & $\delta_{1}$ & $\delta_{2}$ & SOHPIE & NetCoMi & MDiNE & SOHPIE & NetCoMi & MDiNE & SOHPIE & NetCoMi & MDiNE & SOHPIE & NetCoMi & MDiNE \\
 \hline
40 & 20 & 0.05 & 0.05 & \textbf{0.26} & 0.25 & 0.23 & 0.64 & 0.27 & \textbf{0.68} & \textbf{0.68} & 0.26 & 0.31 & \textbf{0.44} & 0.07 & 0.37  \\
& & 0.05 & 0.10 & \textbf{0.34} & 0.33 & 0.27 & 0.64 & 0.28 & \textbf{0.68} & 0.43 & 0.30 & 0.40 & \textbf{0.46} & 0.09 & 0.41 \\
& & 0.05 & 0.20 & \textbf{0.50} & 0.47 & 0.46 & \textbf{0.66} & 0.24 & 0.64 & 0.55 & 0.31 & \textbf{0.60} & \textbf{0.50} & 0.12 & 0.49  \\
& & 0.10 & 0.05 & \textbf{0.35} & 0.34 & \textbf{0.35} & \textbf{0.65} & 0.28 & 0.59 & 0.44 & 0.30 & \textbf{0.48} & 0.46 & 0.09 & \textbf{0.48} \\
& & 0.10 & 0.10 & \textbf{0.42} & 0.40 & 0.35 & \textbf{0.65} & 0.27 & 0.38 & \textbf{0.49} & 0.32 & 0.35 & 0.48 & 0.11 & \textbf{0.49} \\
& & 0.10 & 0.20 & 0.53 & 0.50 & \textbf{0.54} & \textbf{0.67} & 0.24 & 0.50 & \textbf{0.58} & 0.32 & 0.51 & 0.51 & 0.13 & \textbf{0.52} \\
& & 0.20 & 0.05 & \textbf{0.50} & 0.47 & 0.47 & \textbf{0.65} & 0.23 & 0.58 & 0.55 & 0.31 & \textbf{0.58} & 0.50 & 0.12 & \textbf{0.51}  \\
& & 0.20 & 0.10 & 0.53 & 0.51 & \textbf{0.57} & \textbf{0.66} & 0.24 & 0.38 & 0.57 & 0.32 & \textbf{0.60} & \textbf{0.51} & 0.13 & 0.49 \\
& & 0.20 & 0.20 & 0.58 & 0.57 & \textbf{0.59} & 0.69 & 0.22 & \textbf{0.84} & 0.62 & 0.32 & \textbf{0.69} & 0.53 & 0.13 & \textbf{0.58}  \\
& 50 & 0.05 & 0.05 & 0.25 & \textbf{0.26} & \textbf{0.26} & \textbf{0.83} & 0.44 & 0.57 & \textbf{0.38} & 0.31 & 0.35 & 0.34 & 0.11 & \textbf{0.48} \\
& & 0.05 & 0.10 & \textbf{0.34} & \textbf{0.34} & 0.33 & \textbf{0.84} & 0.45 & 0.67 & \textbf{0.48} & 0.37 & 0.42 & 0.40 & 0.15 & \textbf{0.44} \\
& & 0.05 & 0.20 & \textbf{0.50} & 0.48 & 0.49 & \textbf{0.84} & 0.40 & 0.65 & \textbf{0.62} & 0.41 & 0.53 & \textbf{0.50} & 0.20 & \textbf{0.50} \\
& & 0.10 & 0.05 & 0.34 & 0.34 & \textbf{0.36} & \textbf{0.83} & 0.44 & 0.64 & \textbf{0.47} & 0.36 & 0.42 & 0.39 & 0.15 & \textbf{0.45} \\
& & 0.10 & 0.10 & 0.41 & 0.41 & \textbf{0.44} & \textbf{0.84} & 0.48 & 0.63 & \textbf{0.54} & 0.41 & 0.48 & 0.44 & 0.19 & \textbf{0.49} \\
& & 0.10 & 0.20 & \textbf{0.53} & 0.52 & \textbf{0.53} & \textbf{0.84} & 0.43 & 0.69 & \textbf{0.64} & 0.44 & 0.55 & \textbf{0.52} & 0.23 & 0.51\\
& & 0.20 & 0.05 & \textbf{0.49} & 0.48 & \textbf{0.49} & \textbf{0.84} & 0.40 & 0.65 & \textbf{0.61} & 0.41 & 0.50 & \textbf{0.50} & 0.20 & \textbf{0.50} \\
& & 0.20 & 0.10 & 0.53 & 0.51 & \textbf{0.55} & \textbf{0.83} & 0.41 & 0.52 & \textbf{0.64} & 0.43 & 0.45 & \textbf{0.52} & 0.22 & 0.50\\
& & 0.20 & 0.20 & \textbf{0.58} & 0.57 & 0.51 & \textbf{0.84} & 0.40 & 0.55 & \textbf{0.68} & 0.44 & 0.48 & \textbf{0.56} & 0.23 & 0.49 \\
& 200 & 0.05 & 0.05 & 0.25 & \textbf{0.26} & 0.25 & \textbf{0.95} & 0.65 & 0.92 & \textbf{0.39} & 0.36 & \textbf{0.39} & 0.28 & 0.16 & \textbf{0.29}  \\
& & 0.05 & 0.10 & 0.34 & \textbf{0.35} & 0.33 & \textbf{0.95} & 0.70 & 0.92 & \textbf{0.50} & 0.45 & 0.48 & \textbf{0.36} & 0.24 & 0.35 \\
& & 0.05 & 0.20 & \textbf{0.50} & \textbf{0.50} & 0.49 & \textbf{0.95} & 0.64 & 0.93 & \textbf{0.65} & 0.53 & 0.63 & \textbf{0.50} & 0.32 & 0.49 \\
& & 0.10 & 0.05 & 0.34 & \textbf{0.35} & 0.34 & \textbf{0.95} & 0.69 & 0.93 & \textbf{0.50} & 0.44 & 0.49 & \textbf{0.36} & 0.23 & \textbf{0.36}\\
& & 0.10 & 0.10 & 0.41 & \textbf{0.42} & \textbf{0.42} & \textbf{0.95} & 0.72 & \textbf{0.95} & 0.57 & 0.51 & \textbf{0.58} & 0.42 & 0.30 & \textbf{0.43} \\
& & 0.10 & 0.20 & 0.53 & 0.53 & \textbf{0.55} & \textbf{0.95} & 0.68 & \textbf{0.95} & 0.68 & 0.57 & \textbf{0.69} & 0.53 & 0.36 & \textbf{0.54} \\
& & 0.20 & 0.05 & \textbf{0.49} & \textbf{0.49} & \textbf{0.49} & \textbf{0.95} & 0.63 & \textbf{0.95} & \textbf{0.64} & 0.52 & \textbf{0.64} & \textbf{0.49} & 0.31 & \textbf{0.49} \\
& & 0.20 & 0.10 & 0.52 & \textbf{0.53} & \textbf{0.53} & \textbf{0.95} & 0.67 & 0.94 & \textbf{0.67} & 0.57 & \textbf{0.67} & \textbf{0.52} & 0.35 & \textbf{0.52} \\
& & 0.20 & 0.20 & \textbf{0.58} & \textbf{0.58} & 0.57 & \textbf{0.95} & 0.63 & 0.93 & \textbf{0.71} & 0.58 & 0.70 & \textbf{0.57} & 0.37 & \textbf{0.57} \\ 
& 500 & 0.05 & 0.05 & \textbf{0.25} & \textbf{0.25} & 0.24 & 0.98 & 0.78 & \textbf{0.99} & \textbf{0.39} & 0.37 & 0.38 & \textbf{0.26} & 0.20 & 0.24 \\
& & 0.05 & 0.10 & 0.34 & \textbf{0.35} & \textbf{0.35} & 0.97 & 0.82 & \textbf{0.99} & 0.50 & 0.47 & \textbf{0.51} & \textbf{0.35} & 0.28 & \textbf{0.35} \\
& & 0.05 & 0.20 & 0.49 & \textbf{0.50} & \textbf{0.50} & \textbf{0.98} & 0.80 & \textbf{0.98} & \textbf{0.65} & 0.59 & \textbf{0.65} & 0.49 & 0.39 & \textbf{0.50}  \\
& & 0.10 & 0.05 & 0.34 & \textbf{0.35} & 0.33 & 0.98 & 0.82 & \textbf{0.99} & \textbf{0.50} & 0.48 & 0.49 & \textbf{0.35} & 0.28 & 0.34 \\
& & 0.10 & 0.10 & \textbf{0.41} & \textbf{0.41} & 0.40 & \textbf{0.98} & 0.85 & \textbf{0.98} & \textbf{0.57} & 0.54 & \textbf{0.56} & \textbf{0.41} & 0.35 & 0.40 \\
& & 0.10 & 0.20 & 0.52 & \textbf{0.53} & \textbf{0.53} & 0.98 & 0.83 & \textbf{1.00} & \textbf{0.68} & 0.63 & \textbf{0.68} & 0.52 & 0.43 & \textbf{0.53} \\
& & 0.20 & 0.05 & 0.50 & 0.50 & \textbf{0.52} & 0.98 & 0.79 & \textbf{1.00} & 0.65 & 0.59 & \textbf{0.68} & 0.50 & 0.39 & \textbf{0.52}  \\
& & 0.20 & 0.10 & \textbf{0.52} & \textbf{0.52} & 0.51 & 0.98 & 0.84 & \textbf{0.99} & \textbf{0.67} & 0.63 & \textbf{0.67} & \textbf{0.52} & 0.44 & 0.51 \\
& & 0.20 & 0.20 & \textbf{0.58} & 0.57 & \textbf{0.58} & 0.98 & 0.81 & \textbf{1.00} & 0.72 & 0.65 & \textbf{0.73} & 0.57 & 0.47 & \textbf{0.58} \\ \hline
\end{tabular}}{}
}
\end{sidewaystable}
\bigskip

% Table 3 below (Univariable results)
\begin{sidewaystable}
\centering
\caption{\label{MiNW:table3} The simulation results for the case when the network structure does not depend on age covariate. The binary group variable in the univariable regression model (binary group only) using pseudo-value approach is compared with NetCoMi and MDiNE with 1,000 replicates. A random network is generated at each simulation replicate. The best results are highlighted in boldface.}
\makebox[\textwidth][c]{
{\footnotesize%fontsize{5}{7}\selectfont
\begin{tabular}{@{}m{.2cm}m{.3cm}m{.5cm} | m{1.0cm}m{1.0cm}m{1.0cm} |  m{1.0cm}m{1.0cm}m{1.0cm} |  m{1.0cm}m{1.0cm}m{1.0cm} | m{1.0cm}m{1.0cm}m{1.0cm} }
 \hline
 &  &  &  \multicolumn{3}{c|}{Precision} & \multicolumn{3}{c|}{Recall} & \multicolumn{3}{c|}{F1} & \multicolumn{3}{c}{Accuracy} \\
\cline{4-6}\cline{7-9}\cline{10-12}\cline{13-15}
$p$ & $n$ & $\delta$ & SOHPIE & NetCoMi & MDiNE & SOHPIE & NetCoMi & MDiNE & SOHPIE & NetCoMi & MDiNE & SOHPIE & NetCoMi & MDiNE \\
 \hline
20 & 20 & 0.05 & \textbf{0.15} & 0.14 & 0.00 & \textbf{0.67} & 0.15 & 0.00 & 0.26 & \textbf{0.33} & 0.00 & 0.39 & 0.02 & \textbf{0.85} \\
& & 0.10 & 0.27 & 0.29 & \textbf{0.42} & \textbf{0.68} & 0.16 & 0.00 & \textbf{0.38} & 0.32 & 0.28 & 0.43 & 0.04 & \textbf{0.73}  \\
& & 0.20 & \textbf{0.47} & 0.46 & 0.25 & \textbf{0.67} & 0.16 & 0.01 & \textbf{0.53} & 0.31 & 0.24 & 0.49 & 0.07 & \textbf{0.53}   \\ 
& 50 & 0.05 & \textbf{0.14} & \textbf{0.14} & 0.11 & \textbf{0.82} & 0.32 & 0.03 & 0.24 & 0.28 & \textbf{0.42} & 0.27 & 0.04 & \textbf{0.83}  \\
& & 0.10 & \textbf{0.27} & \textbf{0.27} & 0.26 & \textbf{0.82} & 0.33 & 0.04 & \textbf{0.39} & 0.33 & 0.31 & 0.35 & 0.09 & \textbf{0.72}  \\
& & 0.20 & \textbf{0.46} & \textbf{0.46} & 0.42 & \textbf{0.81} & 0.33 & 0.04 & \textbf{0.58} & 0.40 & 0.23 & 0.47 & 0.15 & \textbf{0.53}   \\
& 200 & 0.05 & \textbf{0.14} & \textbf{0.14} & 0.13 & \textbf{0.94} & 0.38 & 0.36 & 0.24 & \textbf{0.28} & 0.27 & 0.19 & 0.05 & \textbf{0.58}   \\
& & 0.10 & \textbf{0.27} & \textbf{0.27} & 0.25 & \textbf{0.93} & 0.39 & 0.36 & \textbf{0.41} & 0.35 & 0.33 & 0.30 & 0.11 & \textbf{0.54}  \\
& & 0.20 & \textbf{0.48} & \textbf{0.48} & 0.44 & \textbf{0.93} & 0.40 & 0.36 & \textbf{0.62} & 0.43 & 0.40 & 0.48 & 0.19 & \textbf{0.49}   \\
& 500 & 0.05 & \textbf{0.14} & \textbf{0.14} & \textbf{0.14} & \textbf{0.97} & 0.52 & 0.67 & 0.24 & \textbf{0.26} & 0.25 & 0.17 & 0.07 & \textbf{0.38}  \\
& & 0.10 & 0.27 & \textbf{0.28} & 0.26 & \textbf{0.97} & 0.54 & 0.65 & \textbf{0.41} & 0.37 & 0.37 & 0.28 & 0.14 & \textbf{0.42}  \\
& & 0.20 & 0.47 & \textbf{0.48} & 0.45 & \textbf{0.96} & 0.54 & 0.64 & \textbf{0.62} & 0.49 & 0.51 & \textbf{0.47} & 0.25 & \textbf{0.47}  \\ \hline
40 & 20 & 0.05 & 0.14 & 0.13 & \textbf{0.20} & 0.61 & 0.18 & \textbf{0.66} & 0.23 & 0.23 & \textbf{0.29} & 0.44 & 0.02 & \textbf{0.48} \\
& & 0.10 & \textbf{0.26} & \textbf{0.26} & \textbf{0.26} & 0.60 & 0.19 & \textbf{0.74} & 0.35 & 0.24 & \textbf{0.40} & \textbf{0.46} & 0.05 & 0.38  \\
& & 0.20 & \textbf{0.46} & 0.45 & 0.44 & 0.60 & 0.18 & \textbf{0.79} & 0.51 & 0.26 & \textbf{0.60} & \textbf{0.49} & 0.08 & 0.46 \\ 
& 50 & 0.05 & \textbf{0.14} & \textbf{0.14} & 0.13 & \textbf{0.84} & 0.25 & 0.73 & \textbf{0.24} & 0.22 & 0.23 & 0.27 & 0.04 & \textbf{0.33}  \\
& & 0.10 & \textbf{0.26} & 0.25 & 0.24 & \textbf{0.82} & 0.24 & 0.70 & \textbf{0.39} & 0.25 & 0.36 & 0.34 & 0.06 & \textbf{0.39}  \\
& & 0.20 & \textbf{0.46} & \textbf{0.46} & 0.42 & \textbf{0.83} & 0.25 & 0.66 & \textbf{0.59} & 0.32 & 0.50 & \textbf{0.48} & 0.11 & 0.47  \\
& 200 & 0.05 & 0.14 & 0.14 & \textbf{0.15} & \textbf{0.94} & 0.32 & \textbf{0.94} & 0.24 & 0.22 & \textbf{0.25} & 0.18 & 0.05 & \textbf{0.19}  \\
& & 0.10 & 0.26 & \textbf{0.27} & \textbf{0.27} & \textbf{0.95} & 0.33 & 0.91 & \textbf{0.41} & 0.30 & \textbf{0.41} & 0.29 & 0.09 & \textbf{0.31}  \\
& & 0.20 & \textbf{0.47} & \textbf{0.47} & \textbf{0.47} & \textbf{0.95} & 0.32 & 0.91 & \textbf{0.62} & 0.37 & 0.61 & \textbf{0.47} & 0.15 & \textbf{0.47} \\
& 500 & 0.05 & 0.14 & 0.14 & \textbf{0.15} & 0.97 & 0.41 & \textbf{1.00} & 0.24 & 0.22 & \textbf{0.26} & \textbf{0.16} & 0.06 & \textbf{0.16} \\
& & 0.10 & \textbf{0.27} & 0.26 & 0.26 & 0.97 & 0.42 & \textbf{0.99} & \textbf{0.41} & 0.32 & \textbf{0.41} & \textbf{0.28} & 0.11 & 0.26 \\
& & 0.20 & \textbf{0.47} & \textbf{0.47} & 0.46 & 0.98 & 0.42 & \textbf{0.99} & \textbf{0.63} & 0.43 & 0.62 & \textbf{0.47} & 0.20 & 0.46 \\ \hline
\end{tabular}}{}
}
\end{sidewaystable}
\bigskip

\newpage

% Turn off if Appendix not used below:
\begin{appendices}

\section{Supplementary Table}

%An appendix contains supplementary information that is not an essential part of the text itself but which may be helpful in providing a more comprehensive understanding of the research problem or it is information that is too cumbersome to be included in the body of the paper.

%%=============================================%%
%% For submissions to Nature Portfolio Journals %%
%% please use the heading ``Extended Data''.   %%
%%=============================================%%

%%=============================================================%%
%% Sample for another appendix section			       %%
%%=============================================================%%

%% \section{Example of another appendix section}\label{secA2}%
%% Appendices may be used for helpful, supporting or essential material that would otherwise 
%% clutter, break up or be distracting to the text. Appendices can consist of sections, figures, 
%% tables and equations etc.

\begin{table}[!htb]
\renewcommand\thetable{S1} 
    \centering
    \setlength{\tabcolsep}{6pt}
\caption{\label{MiNW:tableS1}Comparison of computational time of SOHPIE-DNA with that of NetCoMi and MDiNE. Multivariable setting is depicted for the illustrative purpose. For each sample size, the minimum and maximum computational times among various combinations of effect sizes are selected. Random network is generated at each simulation replicate.}
{\normalsize%fontsize{5}{7}\selectfont
\begin{tabular}{ll  m{2cm}m{2cm}m{2.1cm}  }
%\begin{tabularx}{\textwidth}{p{.5cm} p{.6cm} | p{1.2cm} p{1.2cm} p{1.2cm} | p{1.2cm} p{1.2cm} p{1.2cm} | p{1.2cm} p{1.2cm} p{1.2cm}}
%\begin{tabular}{ c | c | c c c | c c c | c c c  }
 \hline
& & \multicolumn{3}{c}{Minimum-Maximum Time (in hours)}  \\ [0.5ex]
$p$ & $n$  & SOHPIE & NetCoMi & MDiNE  \\
 \hline
20 & 20 & 0.92-1.98 & 0.47-0.58 & 9.00-13.68\\
& 50 & 2.12-3.42 & 0.52-0.80 & 8.17-15.95 \\
& 200 & 9.57-13.85 & 0.43-0.52 & 11.00-45.00 \\
& 500 & 28.40-50.40 & 0.57-0.80 & 11.00-50.63 \\ \hline
40 & 20 & 2.13-3.15 & 0.73-1.52 & 14.67-32.63 \\
& 50 & 3.47-6.32 & 0.8-0.93 & 37.62-58.00  \\
& 200 & 14.97-19.62 & 0.92-1.43 & 180.65-320.28 \\
& 500 & 50.73-80.05  & 1.22-2.78 & 455.47-496.80 \\ \hline
\end{tabular}}{}
\end{table}

\end{appendices}

%%===========================================================================================%%
%% If you are submitting to one of the Nature Portfolio journals, using the eJP submission   %%
%% system, please include the references within the manuscript file itself. You may do this  %%
%% by copying the reference list from your .bbl file, paste it into the main manuscript .tex %%
%% file, and delete the associated \verb+\bibliography+ commands.                            %%
%%===========================================================================================%%
\newpage 

\end{document}